\documentclass[aps,prl,twocolumn,showpacs]{revtex4}
\usepackage{graphicx} 
\usepackage{dcolumn}  
\usepackage{bm}       

\def\ii{{\rm i}}     \def\Eb{{\bf E}}
\def\ee{{\rm e}}   \def\Rb{{\bf R}}  
   \def\Hb{{\bf H}}   \def\kb{{\bf k}}

\def\aE{{\alpha_{\rm E}}}  \def\aM{{\alpha_{\rm M}}}
\def\apE{{\alpha'_{\rm E}}}  \def\apM{{\alpha'_{\rm M}}}

\begin{document}
\title{Site and lattice resonances in metallic hole arrays}
\author{F.~J.~Garc\'{\i}a~de~Abajo,$^{1}$ J.~J. S\'{a}enz,$^2$ I. Campillo,$^3$ and J. S. Dolado,$^3$}
\affiliation{$^1$Centro Mixto CSIC-UPV/EHU and Donostia
International
Physics Center (DIPC) Aptdo. 1072, 20080 San Sebastian, Spain \\
$^2$Departamento de F\'{\i}sica de la Materia Condensada,
Universidad Aut\'{o}noma de Madrid, 28049 Madrid, Spain \\
$^3$LABEIN Centro Tecnol\'{o}gico, NANOC, Cuesta de Olabeaga 16,
48013 Bilbao, Spain}

\date{\today}

\begin{abstract}
A powerful analytical approach is followed to study light
transmission through subwavelength holes drilled in thick
perfect-conductor films, showing that full transmission (100\%) is
attainable in arrays of arbitrarily narrow holes as compared to
the film thickness. The interplay between resonances localized in
individual holes and lattice resonances originating in the array
periodicity reveals new mechanisms of transmission enhancement and
suppression. In particular, localized resonances obtained by
filling the holes with high-index-of-refraction material are
examined and experimentally observed through large enhancement in
the transmission of individual holes.
\end{abstract}

\pacs{42.25.Fx,42.79.Dj,41.20.Jb,73.20.Mf}


\maketitle

Light scattering from subwavelength holes drilled in metals has
been the subject of long-standing interest \cite{B1944} motivated
by phenomena such as extraordinary light transmission
\cite{C1971,M1980,ELG98} that challenges the severe
$(a/\lambda)^4$ cut-off predicted by Bethe \cite{B1944} for the
transmission cross section of single holes of small radius $a$
compared to the wavelength $\lambda$. In particular, 100\%
transmission was predicted to be attainable using lossless metals
\cite{C1971} and confirmed experimentally for quasi-perfect
conductors in the microwave and THz domains \cite{M1980}. Quite
different from perfect-conductors, real metals are capable of
sustaining surface plasmons that were readily recognized to
mediate the interaction among arrayed holes at visible and
near-infrared frequencies \cite{ELG98,MGL01,BMD04}. Furthermore,
the strong correlation of the transmission enhancement with the
lattice periodicity in both of these metallic regimes has prompted
rigorous descriptions of the effect in terms of dynamical
diffraction \cite{T99_02,SVV03,LT04}, which connects directly to
Wood's anomalies \cite{W1935,SVV03}.

Transmission resonances in individual holes
\cite{LDD02,paper069,FLL04} offer an additional handle to achieve
extraordinary effects. These resonances can be triggered by
decorating a hole with a grating around it \cite{LDD02} (similar
to some directional antennas designs \cite{SK94}), by filling the
hole with high-index-of-refraction material \cite{paper069}, or by
changing its shape to induce strong polarization
\cite{FLL04,KES04}. The combination of {\em lattice resonances}
\cite{CE1961_1962} in hole arrays and {\em site resonances} at
specific hole positions can be anticipated to yield interesting
properties in line with recent studies of light reflection on
metal surfaces patterned with nanocavities that support localized
modes \cite{CNB01,trans1}.


In this Letter, we offer a systematics to study the phenomenology
associated to light transmission through subwavelength hole arrays
in perfect-conductor films, which permits us to establish the
existence of full transmission resonances for arbitrarily small
holes in thick metallic screens. Furthermore, individual-hole
resonances are obtained by filling the holes with
high-index-of-refraction materials. This gives rise to enhanced
subwavelength-light transmission, which is demonstrated both
theoretically and experimentally. Finally, the complex scenario
that is presented when transmission resonances of individual holes
are combined with resonances originating in the array periodicity
is elucidated within our analysis.

In his pioneering development, Bethe \cite{B1944} showed that the
scattered far-field from a hole drilled in an infinitely-thin
perfect-conductor screen can be assimilated to that of a magnetic
dipole parallel to the screen plus an electric dipole
perpendicular to it. Subsequent studies supplemented this result
with higher-order multipole corrections \cite{B1954}, and
eventually, with rigorous solutions for arbitrary hole radius and
film thickness \cite{R1987,paper069}. Small holes can be still
represented by induced dipoles in thick screens, as illustrated in
Fig.\ \ref{Fig1}. This allows defining electric (E) and magnetic
(M) polarizabilities both on the same side as the applied field
($\alpha_\nu$, with $\nu=$E,M) and on the opposite side
($\alpha'_\nu$). Flux conservation under arbitrary illumination
leads to an exact optical-theorem type of relationship between
these polarizabilities:
\begin{eqnarray}
  {\rm Im}\{g^\pm_\nu\} = {\rm Im}\{\frac{1}{\alpha_\nu\pm\alpha'_\nu}\} = \frac{-2
  k^3}{3},
\label{OT}
\end{eqnarray}
where $k=2\pi/\lambda$ is the momentum of light in free space. The
remaining real parts of
$g^\pm_\nu=(\alpha_\nu\pm\alpha'_\nu)^{-1}$ are obtained
numerically \cite{R1987,paper069} and represented in Fig.\
\ref{Fig1}b-c for empty holes. Note that ${\rm Re}\{g^+_\nu\}$
diverges for zero thickness, in which case ${\rm
Re}\{\alpha_\nu\}=-{\rm Re}\{\alpha'_\nu\}$ \cite{B1944}, and that
$|{\rm Re}\{\alpha'_{\rm M}\}|\gg |{\rm Re}\{\alpha'_{\rm E}\}|$
in the thick-film limit, dominated by transmission via the
lowest-frequency TE guided mode, that does not create electric
polarization.

When a subwavelength hole is filled with dielectric material of
sufficiently high permittivity $\epsilon$, hole-cavity resonances
can exist thanks to the reduction of the wavelength by a factor of
$\sqrt{\epsilon}$. These resonances give rise to enhanced
transmission \cite{paper069}, as shown in Fig.\ \ref{Fig2}a by
rigorous numerical solution of Maxwell's equations (curves)
\cite{R1987,paper069}. We present experimental evidence of this
hole-resonance behavior in Fig.\ \ref{Fig2}b, which compares the
transmission of microwaves through subwavelength holes filled with
teflon and air. A 5-fold enhancement in the transmission is
observed.

The width of these resonances is dictated by coupling of the
cavity modes to the continuum of light states outside the film.
The resonances are of Fabry-Perot origin \cite{trans22}, but the
transmission line shapes are actually determined from the noted
coupling to the two continua outside the film, as described by
Fano \cite{F1961} (the vanishing of the transmission when
$g^+_{\rm M}=g^-_{\rm M}$, i.e., $\alpha'_M=0$, is a signature of
a Fano resonance; see below). The coupling strength drops rapidly
for large $\epsilon$ due in part to small transmission through the
dielectric-air interface, as predicted by Fresnel's equations. The
larger $\epsilon$, the narrower the resonance, and the higher the
transmission maxima. Incidentally, the normalized transmission
cross-section obtained from our effective dipole model
($16k^4|\apM|^2/3a^2$, symbols in Fig.\ \ref{Fig2}a for
$\epsilon=50$) compares remarkably well with the exact result
(curve) for small $a/\lambda$.


Periodic arrays of sufficiently small and spaced holes can also be
described by perpendicular electric dipoles $p$ and $p'$ and
parallel magnetic dipoles $m$ and $m'$, where primed (unprimed)
quantities are defined on the entry (exit) side of the film as
determined by the incoming light. This is an extension of previous
considerations for thin screens relying on Babinet's principle
\cite{CE1961_1962,trans2}.  We consider first a
unit-electric-field p-polarized plane wave incident on a hole
array with parallel momentum $\kb_\parallel$ along the $x$ axis,
so that the external (incident plus reflected) field in the
absence of the holes has parallel magnetic field $H_y^{\rm ext}=2$
along the $y$ direction and perpendicular electric field $E_z^{\rm
ext}=-2k_\parallel/k$ along $z$. One can write the following set
of multiple-scattering equations for the self-consistent dipoles,
that respond both to the external field and to the field scattered
by the other holes:
\begin{eqnarray}
p &=& \aE (E_z^{\rm ext} + G_z p - H m) + \apE (G_z p' - H m') \nonumber \\
p'&=& \apE (E_z^{\rm ext} + G_z p - H m) + \aE (G_z p' - H m') \nonumber \\
m &=& \aM (H_y^{\rm ext} + G_y m - H p) + \apM (G_y m' - H p') \nonumber \\
m'&=& \apM (H_y^{\rm ext} + G_y m - H p) + \aM (G_y m' + H p'),
\nonumber \nonumber
\end{eqnarray}
where $G_j$ and $H$ describe the induced fields produced at a
given hole by the other holes \cite{CE1961_1962}. Noticing that
the dipoles depend on hole positions $\Rb=(x,y)$ only via phase
factors $\exp(\ii k_\parallel x)$, one finds
\begin{eqnarray}
G_j&=&\sum_{{\bf R}\neq 0} \ee^{-\ii k_\parallel x}
(k^2+\partial_j\partial_j)\frac{{\rm e}^{{\rm i} kR}}{R} \label{GG}\\
H&=&-\ii k\sum_{{\bf R}\neq 0} \ee^{-\ii k_\parallel
x}\partial_x\frac{{\rm e}^{{\rm i} kR}}{R}. \nonumber
\end{eqnarray}
The solution of the above equations can be written
\begin{eqnarray}
p\pm p' &=& -2 [(g_{\rm M}^\pm-G_y) k_\parallel/k+H]/\Delta_\pm \nonumber \\
m\pm m' &=&  2 [(g_{\rm E}^\pm-G_z) +H k_\parallel/k]/\Delta_\pm
\nonumber
\end{eqnarray}
with
\begin{eqnarray}
\Delta_\pm=(g_{\rm E}^\pm-G_z)(g_{\rm M}^\pm-G_y)-H^2, \label{DD}
\end{eqnarray}
from where the zeroth-order transmission of the holey film can be
evaluated as obtained from the far field set up by an infinite 2D
array of dipoles:
\begin{eqnarray}
T=|\frac{2\pi k^2}{A k_z}(m'-p'k_\parallel/k)|^2. \nonumber
\end{eqnarray}
Here, $A$ is the lattice unit-cell area and
$k_z=\sqrt{k^2-k_\parallel^2}$.

Similarly, the transmittance of s-polarized light reduces to
$T=|2\pi k m'/A|^2$, with magnetic dipoles parallel to
$\kb_\parallel$ and no electric dipoles whatsoever ($E^{\rm
ext}_z=0$). More precisely,
\begin{eqnarray}
m\pm m' &=&  \frac{2k_z/k}{g_{\rm M}^\pm-G_x}, \nonumber
\end{eqnarray}
from where one obtains
\begin{eqnarray}
&T&=(\frac{2\pi k_z}{A})^2 |\frac{1}{g^+_{\rm M}-G_x} -
\frac{1}{g^-_{\rm M}-G_x}|^2 \label{Ts} \\
&=& |\frac{1}{1+\frac{\ii A}{2\pi k_z} {\rm Re}\{g^+_{\rm
M}-G_x\}} - \frac{1}{1+\frac{\ii A}{2\pi k_z} {\rm Re}\{g^-_{\rm
M}-G_x\}}|^2. \nonumber
\end{eqnarray}
The last identity in Eq.\ (\ref{Ts}) is derived from Eq.\
(\ref{OT}) and from the exact relation ${\rm Im}\{G_x\}=2\pi
k_z/A-2k^3/3$ for propagating light ($k_\parallel<k$).


The performance of the hole array is dominated by divergences in
the lattice sums when the diffraction orders $(m,n)$ go grazing.
More precisely, for a square lattice of spacing $d$ and for
$\kb_\parallel$ along $x$, the sums $G_j$ go to $+\infty$ as
\begin{eqnarray}
G_j \propto \frac{1}{\sqrt{(k_\parallel+2\pi m/d)^2+(2\pi
n/d)^2-k^2}}. \label{latres}
\end{eqnarray}
In particular, $G_y$ and $G_z$ (p polarization) diverge on the
lowest-frequency side of all grazing diffraction orders, as
illustrated in Fig.\ \ref{Fig3}, whereas $G_x$ (s polarization)
diverges only for $n\neq 0$ (non-straight curves). This entails
different peak structure patterns for s- and p-polarized light
(see Fig.\ \ref{Fig4}).

Interestingly, Eq.\ (\ref{Ts}) predicts 100\% transmission
whenever the condition
\begin{eqnarray}
1+(\frac{A}{2\pi k_z})^2 \, {\rm Re}\{g^+_{\rm M}-G_x\} \, {\rm
Re}\{g^-_{\rm M}-G_x\}=0 \label{T1cond}
\end{eqnarray}
is fulfilled. Eq.\ (\ref{T1cond}) is a second-order algebraic
equation in ${\rm Re}\{G_x\}$ that admits positive real solutions
(one or two depending on the signs of ${\rm Re}\{g^\pm_{\rm M}\}$)
when
\begin{eqnarray}
  \frac{A}{4\pi k_z} |g^+_M-g^-_M| > 1,
\label{T1cond2}
\end{eqnarray}
a condition that can be easily satisfied near $n\neq 0$ grazing
diffraction orders, where $G_x$ can be chosen arbitrarily large
within a narrow range of wavelengths [see Eq.\ (\ref{latres})]. It
should be noted that the difference $g^+_M-g^-_M$ falls off
rapidly to zero when the film thickness $t$ is made much larger
than the hole radius for empty holes (see Fig.\ \ref{Fig1}b).
However, for fixed $t/a$ ratio and angle of incidence, the left
hand side of (\ref{T1cond2}) reduces to a positive real constant
times $\lambda A/a^3$, leading to the conclusion that 100\%
transmission is possible regardless how small the holes are as
compared to the film thickness, provided the separation between
holes (or equivalently $A$) is made sufficiently large.

The interaction between site and lattice resonances is explored in
Fig.\ \ref{Fig4} through the transmittance of square lattices of
holes filled with materials of different permittivity for various
values of the $t/a$ ratio, both for p-polarized and s-polarized
incident light. The dominant features of these plots can be
classified as follows:

{\bf (i)} {\it Full transmission close to lattice resonances as
those of Fig.\ \ref{Fig3}.} In particular under the conditions of
Fig.\ \ref{Fig4}d, one can neglect the first term inside the
squared modulus of Eq.\ (\ref{Ts}) (${\rm Re}\{g^+_{\rm M}\}>>{\rm
Re}\{g^-_{\rm M}\}$, see Fig.\ \ref{Fig1}b), so that 100\%
transmittance maxima come about near $n\neq 0$ grazing diffraction
orders (see Fig.\ \ref{Fig3}) for which one can have ${\rm
Re}\{g^-_{\rm M}\}\approx {\rm Re}\{G_x\}$.

{\bf (ii)} {\it Full transmission close to dispersionless site
resonances.} This is illustrated graphically in the left part of
Fig.\ \ref{Fig5}, which shows that features A and C of Fig.\
\ref{Fig4}j correspond to 100\% transmission at wavelengths where
the condition noted in (i) is fulfilled. The resonant
individual-hole polarizabilities represented in the lower part of
Fig.\ \ref{Fig5} display a typical Lorentzian shape in coincidence
with a transmission maximum for isolated holes (see $\epsilon=100$
curve in Fig.\ \ref{Fig2}a), from which the full transmission
maximum in the lattice is blue-shifted due to inter-hole
interaction described by $G_x$. The density of site resonances
increases dramatically both with film thickness (see Fig.\
\ref{Fig4}l) and with $\epsilon$ (Fabry-Perot-like behavior).

{\bf (iii)} {\it Dispersionless regions of vanishing
transmission.} Eq.\ (\ref{Ts}) predicts $k_\parallel$-independent
vanishing transmission when $g^+_{\rm M}=g^-_{\rm M}$, which is a
property of single holes. This is the case of feature B in Fig.\
\ref{Fig4}j, as illustrated geometrically in the right part of
Fig.\ \ref{Fig5}.

{\bf (iv)} {\it Strong mixing of site and lattice resonances.}
Avoided level crossings are particularly evident in Figs.\
\ref{Fig4}c,g near $k_\parallel=\pi/d$. Non-avoided crossings are
also observed, as well as splitting of full transmission maxima.

{\bf (v)} {\it Film-bound states.} For incident evanescent light
with $k<k_\parallel<2\pi/d-k$, the lattice sums satisfy ${\rm
Im}\{G_j\}=-2k^3/3$ and ${\rm Im}\{H\}=0$. This implies that
$\Delta_\pm$ [Eq.\ (\ref{DD})] is real and can vanish for specific
combinations of $k$ and $k_\parallel$, leading to simultaneous
infinite transmittance and reflectance (evanescent waves do not
propagate energy) in what constitute film-bound resonances, as
recently predicted for related metal structures \cite{PMG04}.

In summary, a simple and powerful formalism has been used to
analyze transmission through hole arrays leading to surprising
results such as 100\% transmission for thick perfect-conductor
films perforated by arbitrarily small holes. Both theoretical and
experimental evidence of single-hole resonances obtained by
filling the holes with large-index-of-refraction material have
been presented, resulting in enhanced transmission through
isolated holes. Finally, filled-hole arrays have been shown to
exhibit a colorful phenomenology, including new types of
suppressed transmission and a complicated interplay between
hole-site resonances and lattice resonances that is explained
within the present approach.



\begin{figure}
\caption{\label{Fig1} {\bf (a)} The field scattered by a
subwavelength hole drilled in a perfect-conductor film in response
to external electric ($E^0$) and magnetic ($H^0$) fields is
equivalent (at a large distance compared to the radius $a$) to
that of effective electric ($p$) and magnetic ($m$) dipoles, which
allow defining polarizabilities ($\alpha_E$ and $\alpha_M$,
respectively) both on the same side as the external fields
($\alpha_\nu$) and on the opposite side ($\alpha'_\nu$). Only the
perpendicular component of the electric field and the parallel
component of the magnetic field induce dipoles. {\bf (b)}-{\bf
(c)} Real part of the hole response functions $g^\pm_\nu$ for
$\epsilon=\mu=1$.}
\end{figure}

\begin{figure}
\caption{\label{Fig2} {\bf (a)} Light transmittance through a
circular hole drilled in a perfect metal film and filled with
dielectric material for different values of the permittivity
$\epsilon$ (see labels). The transmitted power is normalized to
the incoming flux times the hole area. The ratio of the film
thickness to the hole radius is 0.1. {\bf (b)} Ratio of
transmission for $\epsilon=10.2$ (teflon) and $\epsilon=1$ (air)
under the same conditions as in (a): theory (solid curve) vs
experiment (symbols).}
\end{figure}

\begin{figure}
\caption{\label{Fig3} Lattice sum $G_z$ [Eq.\ (\ref{GG})] for a
square lattice of period $d$ as a function of parallel momentum
$k_\parallel$ and wavelength.}
\end{figure}

\begin{figure}
\caption{\label{Fig4} Zeroth order light transmittance through
square arrays of circular holes drilled in perfect-conductor films
as a function of parallel momentum $k_\parallel$ and wavelength
$\lambda$. The ratio of the hole radius to the lattice spacing is
taken as $a/d=0.2$. Different values of the film thickness $t$ and
the dielectric constant inside the holes $\epsilon$ are
considered, as shown by labels. Both p-polarized ($\Hb$ parallel
to the film) and s-polarized ($\Eb$ parallel to the film) incident
light are considered.}
\end{figure}

\begin{figure}
\caption{\label{Fig5} Normal-incidence transmittance (top),
lattice sums and hole response functions (middle), and hole
polarizability (bottom; see Fig.\ \ref{Fig1}a) under the same
conditions as in Fig.\ \ref{Fig4}j ($a/d=0.2$, $t/a=0.1$,
$\epsilon=100$).}
\end{figure}



\begin{thebibliography}{25}
\expandafter\ifx\csname
natexlab\endcsname\relax\def\natexlab#1{#1}\fi
\expandafter\ifx\csname bibnamefont\endcsname\relax
  \def\bibnamefont#1{#1}\fi
\expandafter\ifx\csname bibfnamefont\endcsname\relax
  \def\bibfnamefont#1{#1}\fi
\expandafter\ifx\csname citenamefont\endcsname\relax
  \def\citenamefont#1{#1}\fi
\expandafter\ifx\csname url\endcsname\relax
  \def\url#1{\texttt{#1}}\fi
\expandafter\ifx\csname
urlprefix\endcsname\relax\def\urlprefix{URL }\fi
\providecommand{\bibinfo}[2]{#2}
\providecommand{\eprint}[2][]{\url{#2}}

\bibitem[{\citenamefont{Bethe}(1944)}]{B1944}
\bibinfo{author}{\bibfnamefont{H.~A.} \bibnamefont{Bethe}},
  \bibinfo{journal}{Phys.\ Rev.} \textbf{\bibinfo{volume}{66}},
  \bibinfo{pages}{163} (\bibinfo{year}{1944}).

\bibitem[{\citenamefont{Ebbesen et~al.}(1998)\citenamefont{Ebbesen, Lezec,
  Ghaemi, Thio, and Wolff}}]{ELG98}
\bibinfo{author}{\bibfnamefont{T.~W.} \bibnamefont{Ebbesen}} {\it et al.},
  \bibinfo{journal}{Nature} \textbf{\bibinfo{volume}{391}},
  \bibinfo{pages}{667} (\bibinfo{year}{1998}).

\bibitem[{\citenamefont{Chen}(1971)}]{C1971}
\bibinfo{author}{\bibfnamefont{C.~C.} \bibnamefont{Chen}},
  \bibinfo{journal}{IEEE Trans. Micr. Theory and Tech.}
  \textbf{\bibinfo{volume}{19}}, \bibinfo{pages}{475} (\bibinfo{year}{1971}).

\bibitem[{\citenamefont{{R.~C. McPhedran {\it et al.}}}(1980)}]{M1980}
\bibinfo{author}{\bibnamefont{{R.~C. McPhedran {\it et al.}, in {\it Electromagnetic Theory of Gratings},
  edited by R. Petit}}}
  (\bibinfo{publisher}{Springer}, \bibinfo{address}{Berlin},
  \bibinfo{year}{1980}), p. \bibinfo{pages}{227};
\bibinfo{author}{\bibfnamefont{J.}~\bibnamefont{{G\'{o}mez-Rivas}}} {\it et al.},
  \bibinfo{journal}{Phys.\ Rev.\ B} \textbf{\bibinfo{volume}{68}},
  \bibinfo{pages}{201306(R)} (\bibinfo{year}{2003}).

\bibitem[{\citenamefont{{Mart\'{\i}n-Moreno}
  et~al.}(2001)\citenamefont{{Mart\'{\i}n-Moreno}, {Garc\'{\i}a-Vidal}, Lezec,
  Pellerin, Thio, Pendry, and Ebbesen}}]{MGL01}
\bibinfo{author}{\bibfnamefont{L.}~\bibnamefont{{Mart\'{\i}n-Moreno}}} {\it et al.}, \bibinfo{journal}{Phys.\ Rev.\ Lett.}
  \textbf{\bibinfo{volume}{86}}, \bibinfo{pages}{1114} (\bibinfo{year}{2001}).

\bibitem[{\citenamefont{Barnes et~al.}(2004)\citenamefont{Barnes, Murray,
  Dintinger, Devaux, and Ebbesen}}]{BMD04}
\bibinfo{author}{\bibfnamefont{W.~L.} \bibnamefont{Barnes}} {\it et al.},
  \bibinfo{journal}{Phys.\ Rev.\ Lett.} \textbf{\bibinfo{volume}{92}},
  \bibinfo{pages}{107401} (\bibinfo{year}{2004}).

\bibitem[{\citenamefont{Treacy}(1999)}]{T99_02}
\bibinfo{author}{\bibfnamefont{M.~M.~J.} \bibnamefont{Treacy}},
  \bibinfo{journal}{Appl.\ Phys.\ Lett.} \textbf{\bibinfo{volume}{75}},
  \bibinfo{pages}{606} (\bibinfo{year}{1999});
  \bibinfo{journal}{Phys.\ Rev.\ B} \textbf{\bibinfo{volume}{66}},
  \bibinfo{pages}{195105} (\bibinfo{year}{2002}).

\bibitem[{\citenamefont{Sarrazin et~al.}(2003)\citenamefont{Sarrazin, Vigneron,
  and Vigoureux}}]{SVV03}
\bibinfo{author}{\bibfnamefont{M.}~\bibnamefont{Sarrazin}} {\it et al.}, \bibinfo{journal}{Phys.\ Rev.\ B}
  \textbf{\bibinfo{volume}{67}}, \bibinfo{pages}{085415}
  (\bibinfo{year}{2003}).

\bibitem[{\citenamefont{Lezec and Thio}(2004)}]{LT04}
\bibinfo{author}{\bibfnamefont{H.~J.} \bibnamefont{Lezec}} \bibnamefont{and}
  \bibinfo{author}{\bibfnamefont{T.}~\bibnamefont{Thio}},
  \bibinfo{journal}{Opt.\ Express} \textbf{\bibinfo{volume}{12}},
  \bibinfo{pages}{3629} (\bibinfo{year}{2004}).

\bibitem[{\citenamefont{Wood}(1935)}]{W1935}
\bibinfo{author}{\bibfnamefont{R.~W.} \bibnamefont{Wood}},
  \bibinfo{journal}{Phys.\ Rev.} \textbf{\bibinfo{volume}{48}},
  \bibinfo{pages}{928} (\bibinfo{year}{1935}).

\bibitem[{\citenamefont{Lezec et~al.}(2002)\citenamefont{Lezec, Degiron,
  Devaux, Linke, Mart\'{\i}n-Moreno, {Garc\'{\i}a-Vidal}, and Ebbesen}}]{LDD02}
\bibinfo{author}{\bibfnamefont{H.~J.} \bibnamefont{Lezec}} {\it et al.}, \bibinfo{journal}{Science}
  \textbf{\bibinfo{volume}{297}}, \bibinfo{pages}{820} (\bibinfo{year}{2002}).

\bibitem[{\citenamefont{{Garc\'{\i}a de Abajo}}(2002)}]{paper069}
\bibinfo{author}{\bibfnamefont{F.~J.} \bibnamefont{{Garc\'{\i}a de Abajo}}},
  \bibinfo{journal}{Optics Express} \textbf{\bibinfo{volume}{10}},
  \bibinfo{pages}{1475} (\bibinfo{year}{2002}).

\bibitem[{\citenamefont{Falcone et~al.}(2004)\citenamefont{Falcone, Lopetegi,
  Laso, Baena, Bonache, Beruete, {Marqu\'{e}s}, {Mart\'{\i}n}, and
  Sorolla}}]{FLL04}
\bibinfo{author}{\bibfnamefont{F.}~\bibnamefont{Falcone}} {\it et al.},
  \bibinfo{journal}{Phys.\ Rev.\ Lett.} \textbf{\bibinfo{volume}{93}},
  \bibinfo{pages}{197401} (\bibinfo{year}{2004}).

\bibitem[{\citenamefont{Shafai and Kishk}(1994)}]{SK94}
\bibinfo{author}{\bibfnamefont{L.}~\bibnamefont{Shafai}} \bibnamefont{and}
  \bibinfo{author}{\bibfnamefont{A.~A.} \bibnamefont{Kishk}},
  {\em Microwave Horns and Feeds}
  (\bibinfo{publisher}{IEEE Press}, \bibinfo{address}{Berlin},
  \bibinfo{year}{1994}), p. \bibinfo{pages}{227}.

\bibitem[{\citenamefont{{Klein Koerkamp} et~al.}(2004)\citenamefont{{Klein
  Koerkamp}, Enoch, Segerink, {van Hulst}, and Kuipers}}]{KES04}
\bibinfo{author}{\bibfnamefont{K.~J.} \bibnamefont{{Klein Koerkamp}}} {\it et al.},
  \bibinfo{journal}{Phys.\ Rev.\ Lett.} \textbf{\bibinfo{volume}{92}},
  \bibinfo{pages}{183901} (\bibinfo{year}{2004}).

\bibitem[{\citenamefont{Collin and Eggimann}(1961)}]{CE1961_1962}
\bibinfo{author}{\bibfnamefont{R.~E.} \bibnamefont{Collin}} \bibnamefont{and}
  \bibinfo{author}{\bibfnamefont{W.~H.} \bibnamefont{Eggimann}},
  \bibinfo{journal}{IRE Trans. Micr. Theory and Tech.}
  \textbf{\bibinfo{volume}{10}}, \bibinfo{pages}{110}
  (\bibinfo{year}{1961});
  \bibinfo{author}{\bibfnamefont{W.~H.} \bibnamefont{Eggimann}} \bibnamefont{and}
  \bibinfo{author}{\bibfnamefont{R.~E.} \bibnamefont{Collin}},
  \bibinfo{journal}{IRE Trans. Micr. Theory and Tech.}
  \textbf{\bibinfo{volume}{10}}, \bibinfo{pages}{528} (\bibinfo{year}{1962}).

\bibitem[{\citenamefont{Coyle et~al.}(2001)\citenamefont{Coyle, Netti,
  Baumberg, Ghanem, Birkin, Bartlett, and Whittaker}}]{CNB01}
\bibinfo{author}{\bibfnamefont{S.}~\bibnamefont{Coyle}} {\it et al.}, \bibinfo{journal}{Phys.\ Rev.\ Lett.}
  \textbf{\bibinfo{volume}{87}}, \bibinfo{pages}{176801}
  (\bibinfo{year}{2001}).

\bibitem[{tra({\natexlab{a}})}]{trans1}
\bibinfo{note}{T.~V. Teperik {\em et al.}, to be published.}

\bibitem[{\citenamefont{Bouwkamp}(1954)}]{B1954}
\bibinfo{author}{\bibfnamefont{C.~J.} \bibnamefont{Bouwkamp}},
  \bibinfo{journal}{Reports on Progress in Physics}
  \textbf{\bibinfo{volume}{XVIII}}, \bibinfo{pages}{35} (\bibinfo{year}{1954}).

\bibitem[{\citenamefont{Roberts}(1987)}]{R1987}
\bibinfo{author}{\bibfnamefont{A.}~\bibnamefont{Roberts}}, \bibinfo{journal}{J.
  Opt. Soc. Am. A} \textbf{\bibinfo{volume}{4}}, \bibinfo{pages}{1970}
  (\bibinfo{year}{1987}).

\bibitem[{tra({\natexlab{b}})}]{trans22}
\bibinfo{note}{F.~J. Garc\'{\i}a-Vidal {\it et al.}, cond-mat/0503709.}

\bibitem[{tra({\natexlab{b}})}]{F1961}
\bibinfo{note}{U. Fano, Phys.\ Rev. {\bf 124}, 1866 (1961).}

\bibitem[{tra({\natexlab{b}})}]{trans2}
\bibinfo{note}{F.~J. Garc\'{\i}a de Abajo {\it et al.}, to be published.}

\bibitem[{\citenamefont{Pendry et~al.}(2004)\citenamefont{Pendry,
  Mart\'{\i}n-Moreno, and {Garc\'{\i}a-Vidal}}}]{PMG04}
\bibinfo{author}{\bibfnamefont{J.~B.} \bibnamefont{Pendry}} {\it et al.}, \bibinfo{journal}{Science}
  \textbf{\bibinfo{volume}{305}}, \bibinfo{pages}{847} (\bibinfo{year}{2004}).

\end{thebibliography}

\end{document}